\documentclass[12pt,a4paper]{article}
\usepackage{graphicx}
\begin{document}

\title{\bf{The use of a syncytium model of the crystalline lens of the 
human eye to study the light flashes seen by astronauts}}
\author{}
\date{}
\maketitle

\vspace{-1. cm}
\centerline{Giampietro Nurzia$^{1}$, Renato Scrimaglio$^{a}$, 
Bruno Spataro$^{b}$, Francesco Zirilli$^{c}$}
\vspace{0.5 cm}

\small{\it
\centerline{$^{a}$ Dip. di Fisica, Univ. de L'Aquila,
Via Vetoio (Coppito 1), 67010 Coppito (L'Aquila), Italy.}

\centerline{$^{b}$ INFN, Laboratori Nazionali di Frascati, 
P.O. Box 13, 00044 Frascati (Roma), Italy.}

\centerline{$^{c}$ Dip. di Matematica ``G. Castelnuovo'', 
Univ. di Roma ``La Sapienza'', 00185 Roma, Italy.}

\vspace{1. cm}
\rm
Syncytium model to study the light flashes.

\vspace{1. cm}
$^{1}$ Author to whom correspondence should be addressed at: 

Dipartimento di Fisica, Universit\`a de L'Aquila, 
Via Vetoio (Coppito 1), 67010 Coppito (L'Aquila), Italy;

Phone: +39 0862 433043

Fax: +39 0862 433043

E-mail: Giampietro.Nurzia@aquila.infn.it
\newpage

\normalsize
\vspace{1 cm}
\bf{\centerline{Abstract}}
\vspace{0.5 cm}
\rm

Nurzia, G., Scrimaglio, R., Spataro, B., Zirilli, F. The use of a 
syncytium model of the crystalline lens of the human eye to study the 
light flashes seen by astronauts. \it{Radiat. Res.}

\vspace{0.5 cm}
\rm
A syncytium model to study some electrical properties of the eye is 
proposed in the attempt to explain the phenomenon of anomalous Light 
Flashes (LF) perceived by astronauts in orbit. The crystalline lens
is modelled as an ellipsoidal syncytium having a variable relative 
dielectric constant. The mathematical model proposed is given by a 
boundary value problem for a system of two coupled elliptic partial 
differential equations in two unknowns. We use a numerical method to 
compute an approximate solution of this mathematical model and we show 
some numerical results that provide a possible (qualitative) explanation 
of the observed LF phenomenon. In particular, we calculate the energy 
lost in the syncytium by a cosmic charged particle that goes through 
the syncytium and compare the results with those obtained using the 
Geant 3.21 simulation program. We study the interaction 
antimatter-syncytium. We use the Creme96 computer program to evaluate 
the cosmic ray fluxes encountered by the International Space Station. 
\newpage

\section*{\small{\centerline{1 INTRODUZIONE}}}

In the next few years the average time spent on the International 
Space Station by human beings will substantially increase. For this 
reason the safety of human life in the space environment is crucial. 
There is a need to study the effects of cosmic rays particles on the 
human body and particularly on the functionality of the Central Nervous 
System (CNS). 

The visual system has been chosen to ``probe'' the CNS because it is 
particularly sensitive to space environment. In missions Apollo
11 through 17, Skylab 4, Apollo-Soyuz, Mir, Iss, the
astronauts, after some minutes of dark adaptation, observed brief
flashes of white light (LF phenomenon) with the shape of thin or thick 
streaks, single or multiple dots, clouds, etc. \it{(1-3)}. \rm

The specific mechanism of the interaction between ionizing particles and 
the visual system remains uncertain. In order to evaluate
the LF phenomenon it is necessary the simultaneous determination of 
time, nature, energy and trajectory of the particle passing through the
cosmonaut eyes, as well as the cosmonaut LF perception time. Some 
previous experiments are described in \it{(4-6)}. \rm 

A future experiment, named ALTEA \it{(7-10)}\rm, will be activated on 
the International Space Station in 2005. The ALTEA project is aimed at 
the study of the transient and long term effects of cosmic particles on 
the astronaut cerebral functions. It has been funded by the Italian Space 
Agency (ASI) and by the italian National Institute for Nuclear Physics 
(INFN) and ``Highly recommended'' by the European Space Agency (ESA). The
experiment is an international and multidisciplinary collaboration.

The basic instrumentation is composed by a series of active particles 
telescopes, an ElectroEncephaloGrapher (EEG) and a visual stimulator, 
arranged in a helmet shaped device. This instrumentation is able to 
measure simultaneously the dynamics of the functional status of the 
visual system, the cortical electrophysiological activity, and the 
passage through the brain of those particles whose energy is included in
a predetermined window. The three basic instruments can be used 
separately or in any combination, permitting several different 
experiments.

In this paper we analyze a mathematical model able to describe some 
electrical properties of the eye. It is based on a mathematical model
of syncytial tissues, that is tissues where many cells are electrically 
coupled one to the other and to an extracellular medium. We note that 
multicellular syncytia are used to model important tissues such as, for 
example, the eye lens \it{(11-14)}\rm. We use the model of syncytial 
tissues presented in \it{(12), (14)} \rm to suggest a mathematical 
explanation of the LF phenomenon. We note that the eye lens is only a 
part of the eye and that in the scientific literature more sophisticated 
models of the eye exist, see for example \it{(15)}\rm. Finally we have 
pointed out the sensitivity of the electrical behaviour of the proposed 
syncytium model respect to the direction of the particle passing through 
the astronaut visual system. In particular, we have calculated the energy
lost in the syncytium by a cosmic charged particle going through the 
syncytium as a function of the direction of motion of the particle and we
have compared the results obtained with the syncytium model with those 
obtained using the Geant 3.21 simulation program.

In section 2 we describe a mathematical model of a syncitial tissue that 
describes some electrical properties of the crystalline lens and a 
numerical method to approximate the solution of the model presented. In 
section 3 we describe the relative dielectric constant of the crystalline
lens. In section 4 we show some numerical results obtained from the 
numerical solution of the model that could provide a qualitative 
explanation of the LF phenomenon. In section 5 we describe the 
interaction between antimatter and syncytium. In section 6 we show the 
cosmic ray fluxes within the Iss obtained using the Creme96 program. In 
section 7 we describe the Geant simulation of the phenomenon under 
scrutiny and compare the results obtained with Geant 3.21 with those 
obtained using the syncytium model. In section 8 we calculate the 
adsorbed and equivalent energy doses in the crystalline lens due to 
cosmic radiation. In section 9 some simple conclusions are drawn.

\section*{\small{2 THE SYNCYTIUM MODEL AND THE FINITE DIFFERENCE 
APPROXIMATION}}

Let us introduce some notations. Let \bf{R} \rm be the set of real 
numbers, $n$ be a positive integer and \bf{$\mbox{R}^{n}$} \rm be the 
$n$-dimensional real Euclidean space. Let 
$\underline{x} \in$ \bf{$\mbox{R}^{n}$} \rm be a generic vector. Let 
\bf{C} \rm be the set of complex numbers. Let $z \in$ \bf{C}, \rm we 
denote with $Re(z)$ the real part of $z$ and with $Im(z)$ the imaginary 
part of $z$. 

Let us consider an ellipsoid of rotation
$D = \{ \underline{x} = {(x, y, z)}^{t} \in$ \bf{$\mbox{R}^{3}$}\rm: 
${x^{2} \over b^{2}} + {y^{2} \over b^{2}} + {z^{2} \over a^{2}} 
\leq 1\}$ filled with a syncytial tissue. Let $\partial D$ be the 
boundary of $D$. We choose the eccentricity of the ellipsoid 
$e = \sqrt{3}/2$. Then we have $b = a \sqrt{1 - e^{2}} = a/2$. Let 
$\epsilon_{r}(\underline{x})$ be the relative dielectric constant in 
$\underline{x} \in {D}$. Let us apply on a point 
$\underline{x}_{I} \in \partial D$ a time harmonic electric current 
having modulus proportional to $I$, direction $\underline{v}_{I}$ and 
frequency $f$. Let $R_{i} \ge 0$ be the resistivity of the intracellular 
medium, $R_{e} \ge 0$ be the resistivity of the extracellular medium, 
$Y_{m} = G_{m} + i 2 \pi f C_{m} \in$ \bf{C} \rm be the specific 
admittance of the cell membrane, that is the membrane that separates 
the intracellular medium from the extracellular medium. We note that 
$Y_{m}$ depends on $f$, but we suppose $R_{i}$, $R_{e}$, $Y_{m}$ to be
independent of the space variables $\underline{x} \in D$. Let 
$\alpha_{m} \in$ \bf{R} \rm be the fraction of the volume occupied by 
the cell membrane per unit volume of tissue. As a consequence of the 
application of the electric current described above to the syncytium we
have the generation of two different electric potentials: one in the 
intracellular compartment, the other in the extracellular compartment. 
These potentials can be seen as two complex functions having the same 
support $D$. Let $U^{(e)}(\underline{x})$, $U^{(i)}(\underline{x})$, 
$\underline{x} \in D$ be the electric potentials in the intracellular 
and extracellular compartments respectively, then we have \it{(12)}\rm:
 
\begin{eqnarray}
\nabla \cdot [\epsilon_{r}(\underline{x}) \nabla U^{(e)}(\underline{x})]
+ R_{e} \alpha_{m} Y_{m} [U^{(i)}(\underline{x}) - 
U^{(e)}(\underline{x})] = 0,
\qquad \underline{x} \in D,
\label{eq1}
\end{eqnarray}
\vspace{-1. cm}
\begin{eqnarray}
\nabla \cdot [\epsilon_{r}(\underline{x}) \nabla U^{(i)}(\underline{x})]
+ R_{i} \alpha_{m} Y_{m} [U^{(e)}(\underline{x}) - 
U^{(i)}(\underline{x})] = -I R_{i} {{\partial \delta} \over 
{\partial \underline{v}_{I}}} (\underline{x} - \underline{x}_{I}),
\qquad \underline{x} \in D,
\label{eq2}
\end{eqnarray}
\vspace{-1. cm}
\begin{eqnarray}
U^{(e)}(\underline{x}) = 0, \qquad \underline{x} \in \partial D,
\label{eq3}
\end{eqnarray}
\vspace{-1. cm}
\begin{eqnarray}
{1 \over R_{i}} {{\partial U^{(i)}(\underline{x})} \over 
{\partial \underline{\hat{n}}}} + Y_{s} U^{(i)}(\underline{x}) = 0, 
\qquad \underline{x} \in \partial D, 
\label{eq4}
\end{eqnarray}
where $\nabla$ denotes the gradient operator, $\cdot$ denotes the scalar
product, $\delta(\underline{x})$ denotes the Dirac delta and 
$\underline{\hat{n}}(\underline{x})$ is the outward unit normal vector to
$\partial D$ in $\underline{x} \in \partial D$. Note that the boundary 
condition given in eq.(\ref{eq3}) states that the electric current can 
flow from the extracellular medium through the outer membrane located on
$\partial D$ with admittance equal to zero; eq.(\ref{eq4}) states that 
the electric current can flow from the intracellular medium through the 
outer membrane located on $\partial D$ with admittance 
$Y_{s} = G_{s} + i 2 \pi f C_{s} \in \bf{C}$\rm, this admittance depends 
on $f$, but it is supposed to be constant with respect to the space 
variables $\underline{x} \in \partial D$. The term on the right hand side
of eq.(\ref{eq2}) represents the application of the current on the 
tissue, this term represents the effect of the charged particles passing 
through the astronauts visual system and the costant $I$ that appears in
(\ref{eq2}) is proportional to the charge of the particles, 
$\underline{v}_{I}$ represents the direction of motion of these 
particles. This mathematical model is similar to the model derived in 
\it{(14)}\rm, so that we omit its derivation and we suggest to look at 
\it{(12), (14)}\rm, for a detailed explanation of the elementary physics 
that explains the model. 
 
The boundary value problem (\ref{eq1}), (\ref{eq2}), (\ref{eq3}), 
(\ref{eq4}) has a unique solution pair $U^{(e)}(\underline{x})$, 
$U^{(i)}(\underline{x})$, $\underline{x} \in D$. However this solution 
can not be computed explicitely and an approximation method must be used
to evaluate $U^{(e)}(\underline{x})$, $U^{(i)}(\underline{x})$ in $D$. 
In particular for the computation of an approximation of 
$U^{(e)}(\underline{x})$, $U^{(i)}(\underline{x})$ in $D$ we have 
rewritten problem (\ref{eq1}), (\ref{eq2}), (\ref{eq3}), (\ref{eq4}) in 
spherical coordinates and we have approximated the solution using the 
finite difference method. Let $\rho \in [0, a]$, $\theta \in [0, \pi]$, 
$\varphi \in [0, 2 \pi)$ be the spherical coordinates. Let $N_{\rho}$, 
$N_{\theta}$, $N_{\varphi} > 1$ be the number of points of the uniform 
discretization grid in spherical coordinates used in the finite 
difference method along the coordinates $\rho$, $\theta$, $\varphi$ 
respectively, then from problem (\ref{eq1}), (\ref{eq2}), (\ref{eq3}), 
(\ref{eq4}) we obtain a linear system of 
$(2 N_{\rho} - 1)(N_{\theta} - 1)N_{\varphi}$ equations in 
$(2 N_{\rho} - 1)(N_{\theta} - 1)N_{\varphi}$ unknowns. 

In the numerical experience described here we have computed the solution 
of this linear system using the biconjugate gradient method. See 
\it{(16)}\rm, for a description of the method. The components of the 
vector solution of this linear system are an approximation of the 
functions $U^{(e)}(\underline{x})$, $U^{(i)}(\underline{x})$ on the 
previously described grid points.

The values of the parameters appearing in eqs.(\ref{eq1}), (\ref{eq2}), 
(\ref{eq3}), (\ref{eq4}) are shown in Table \ref{tab1}: 

\begin{table*}[h]
\vspace{1 cm}
\begin{center}
\begin{tabular}{|c|c|}
\hline
$a$  &  1.6 mm  \\ 
\hline
$R_{i}$  &  $6.25 \cdot 10^{3} \ \Omega \mbox{mm}$  \\
\hline
$R_{e}$  &  $4.85 \cdot 10^{5} \ \Omega \mbox{mm}$  \\ 
\hline
$G_{m}$  &  $4.38 \cdot 10^{-9} \ \Omega^{-1} \mbox{mm}^{-2}$  \\
\hline
$C_{m}$  &  $0.79 \cdot 10^{-8} \ \mbox{F} \mbox{mm}^{-2}$  \\
\hline
$G_{s}$  &  $2.14 \cdot 10^{-6} \ \Omega^{-1} \mbox{mm}^{-2}$  \\
\hline
$C_{s}$  &  $9.75 \cdot 10^{-8} \ \mbox{F} \mbox{mm}^{-2}$  \\
\hline
$\alpha_{m}$  &  $6 \cdot 10^{2} \ \mbox{mm}^{-1}$  \\
\hline
\end{tabular}
\end{center}
\caption{\it Electrical and morphological parameters of the crystalline 
lens of the frog eye.}
\label{tab1}
\end{table*}

These values are taken from \it{(12)} \rm and they are relative to the 
frog eye. This is a starting point. The study of the problem with 
parameters relative to the human eye will probably give results 
qualitatively similar to those obtained with the data of Table 
\ref{tab1}. However the values of the human eye parameters are not 
available to us at the moment. Moreover we have chosen 
$\underline{x}_{I} = {(0, 0, 1.6)}^{t}$ mm and 
$N_{\rho} = N_{\theta} = N_{\varphi} = 30$. In the next sections the 
values of the electric current, frequency and relative dielectric 
constant used in our work are shown.

\section*{\small{\centerline{3 THE RELATIVE DIELECTRIC COSTANT}}}

In the macroscopic approach the biological tissues are generally 
considered as media that interact with the electric field induced by the 
external environment in two different ways: (1) generating electric 
currents of conduction that increase with the conductivity of tissues;
(2) producing polarization effects that depend on the local dielectric 
constant. For this reason, from an electromagnetic point of view, 
biological tissues can be considered as dielectric media able to store 
and dissipate the energy of the electromagnetic fields involved. 
According to electromagnetic theory the physical quantity that 
characterizes these mechanisms is the complex relative dielectric 
constant. The real part of this constant takes into account the temporary
storage of energy in the medium, while the imaginary part, depending on 
the conductivity $\sigma$, is responsible for the dissipation of the 
electromagnetic energy. 

In this work we assume the crystalline lens of the frog eye to be a 
perfect dielectric ($\sigma = 0$), so that the relative dielectric 
costant $\epsilon_{r}$ has not imaginary part \rm and is the square of 
the refraction index. The crystalline has the shape of a thin biconvex 
lens, is constituted by a very transparent and very elastic substance 
and has a concentric shell structure. Since the crystalline lens has 
variable density, $\epsilon_{r}$ is variable too. 

Applying the finite difference method, we obtain a series of ellipsoidal
shells with constant eccentricity equal to $e$. Let $\rho$ to be the 
distance from the centre of the ellipsoid $D$ (origin of the coordinates
system) calculated along $a$ semiaxis. A shell is determined by a value 
of the variable $\rho$. Then we suppose $\epsilon_{r}$ to be costant 
whithin every shell and given by:

\begin{eqnarray}
\epsilon_{r} = 
\epsilon_{c} - {\rho \over a} (\epsilon_{c} - \epsilon_{s}),
\label{eq5}
\end{eqnarray}
where $\epsilon_{c} = 1.98$ and $\epsilon_{s} = 1.89$ are the values in 
the centre $(\rho = 0)$ and on the boundary of the ellipsoid (determined 
by the shell having $\rho = a$) respectively. The variation of 
$\epsilon_{r}$ between the centre and the boundary of the ellipsoid is 
4.76\%. Figure \ref{fig1} shows the relative dielectric costant vs. 
$\rho$.

\begin{figure*}[htbp]
\centerline{\resizebox{9cm}{9cm}{
\includegraphics*[width=10cm]{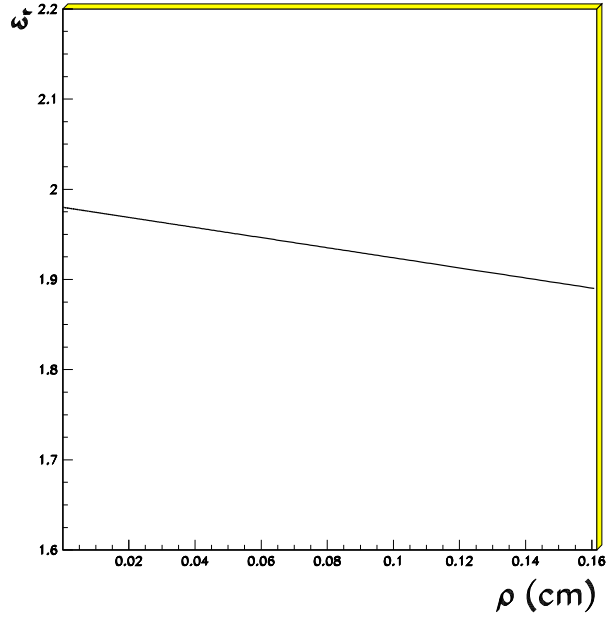}}}
\caption{\it{Relative dielectric costant of the eye vs. $\rho$.}} 
\label{fig1}
\end{figure*}

Let $l$ to be the index corresponding to a value of the $\rho$ variable.
Then $l$ determines the values of the semiaxes of the $l^{th}$shell, 
$a_{l}$ and $b_{l}$. In this way the distance between a point on shell 
and the origin is: 

\begin{eqnarray}
r_{l} = r_{l}(\theta) = {{a_{l} b_{l} \sqrt{1 + \tan^{2}{\theta}}} \over 
{\sqrt{b_{l}^{2} + a_{l}^{2} \tan^{2}{\theta}}}}, \quad 
\theta \in [0, \pi].
\label{eq5b}
\end{eqnarray}
The variable that enters in the equations of the electric potentials is 
$r_{l}$. In the case of the spherical syncytium \it{(14)} \rm we have 
$r_{l} = a_{l}$ when $\theta \in [0, \pi]$. 

A theory describing the ocular lens as a radially nonuniform spherical 
syncytiym is proposed, solved and described as a simple equivalent 
circuit in \it{(13)}\rm. In this paper the syncytium consists of a 
nucleus with one effective intracellular resistivity surrounded by a 
cortex with another resistivity.

\section*{\small{4 THE ELECTRIC POTENTIAL AND THE ENERGY LOST IN THE 
SYNCYTIUM}}

We study the electric potentials of the syncytium with variable density 
as a function of the incidence direction of the cosmic charged particle.
The incidence point $\underline{x}_{I} = {(0, 0, z)}^{t} \in \partial D$ 
is located on the North pole of the ellipsoidal syncytium.

Let $\alpha$ be the incidence angle on the crystalline lens measured 
respect to the positive $z$ axis. When $\alpha = 90^{\circ}$ the 
particle direction of motion is tangent to the ellipsoid and the 
interaction effect is minimum. When $\alpha = 180^{\circ}$ the particle 
direction of motion is along the semiaxis of length $a$ and the 
interaction effect is maximum. 

Figure \ref{fig2} shows the electrical behavior of the syncytium model 
respect to the choice of three different directions for the electric 
current when $f = 3$ Hz and $I = 7 \ \mu \mbox{A}$. This value of the 
electric current comes from the comparison  with the Geant 3.21 
simulation if we imagine that the incident cosmic particle is a proton 
(see section 7, Table \ref{tab2}). In Figure \ref{fig2} we can see the 
shining effect when the electric potentials assume high values.

\begin{figure*}[htbp]
\centerline{\resizebox{15cm}{15cm}{
\includegraphics*[width=10cm]{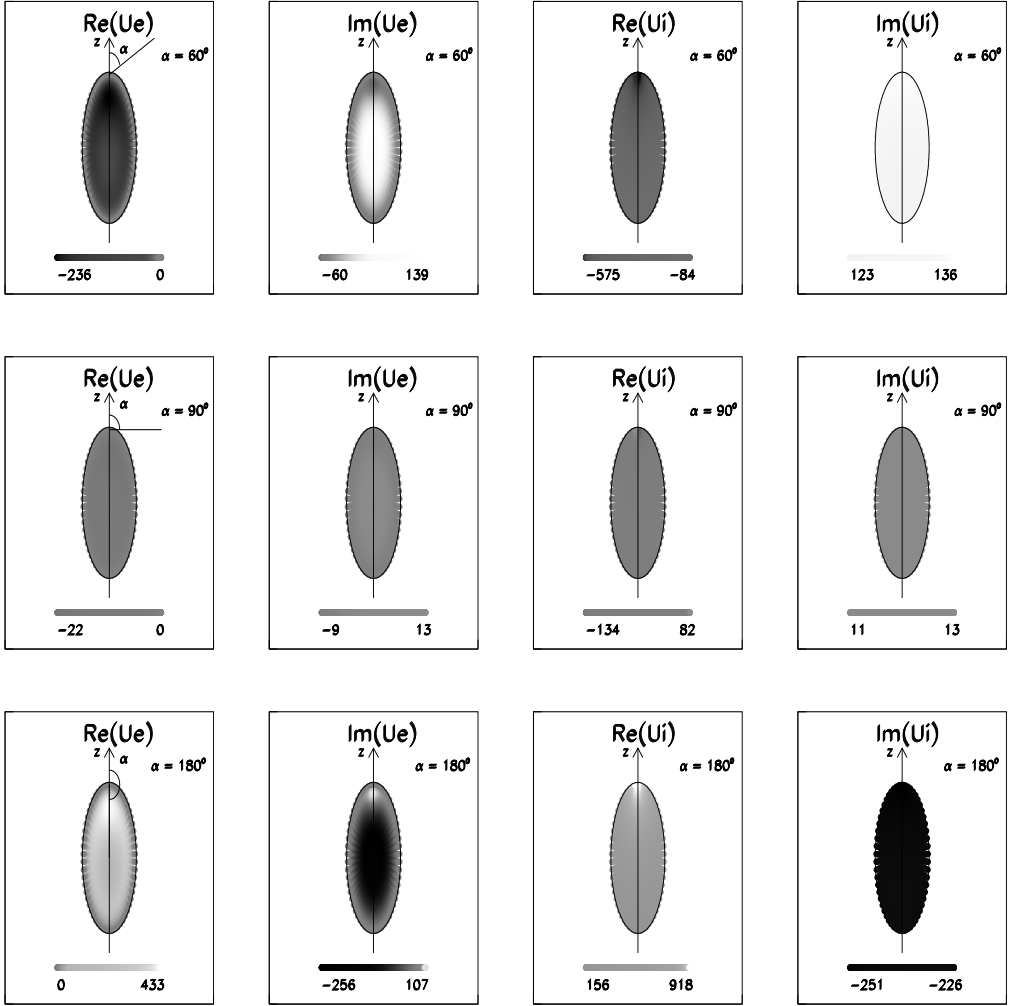}}}
\caption{\it{The electric potentials $U^{(e)}(\underline{x})$, $U^{(i)}(\underline{x})$ in the plane 
$x = 0$. We have chosen the directions $\alpha = 60^{\circ}$, 
$90^{\circ}$ and $180^{\circ}$, the frequency $f = 3$ Hz and the 
electric current $I = 7$ $\mu$A. In order to do a comparison between 
the potentials corresponding to different incidence directions, in each 
column the values of the potentials are normalized to the absolute 
maximum and minimun values. In this way the resulting values are 
dimensionless and are represented on a grey scale between 0 (dark grey) 
and 1 (bright grey). The bars are drawn with the normalized values. Then
in each column the same linear grey-scale is used, but for figures on 
different columns different scales are used. But the numbers written 
below the bars are the real maximum and minimum values of the electric 
potentials and they are expressed in millivolts. We can see that when 
these values are high a shining effect exists.}}
\label{fig2}
\end{figure*}

We have calculated the gradients of the electric potentials using the 
finite difference method in spherical coordinates. The energy lost by a 
cosmic charged particle in the intracellular and extracellular 
compartments of the syncytium is respectively:  

\begin{eqnarray}
\Delta E^{(i)} = {\epsilon_{0} \over 2} \int_{V_{e}} 
\epsilon_{r} {\mid \nabla U^{(i)} \mid}^{2} dV_{e},
\label{eq6}
\end{eqnarray}
and
  
\begin{eqnarray}
\Delta E^{(e)} = {\epsilon_{0} \over 2} \int_{V_{e}} 
\epsilon_{r} {\mid \nabla U^{(e)} \mid}^{2} dV_{e},
\label{eq7}
\end{eqnarray}
where $\epsilon_{0} = 8.854 \cdot 10^{-12} \
{\mbox{C}^{2} \over {\mbox{N} \mbox{m}^{2}}}$ (SI units) is the vacuum 
dielectric constant and $V_{e}$ is the volume of the ellipsoid $D$.

The total energy lost in the syncytium is: 

\begin{eqnarray}
\Delta E^{(t)} = \Delta E^{(i)} + \Delta E^{(e)}. 
\label{eq8}
\end{eqnarray}

Figure \ref{fig3} shows the behavior of the energy lost in the
intracellular compartment (I), in the extracellular compartment (E) and
in the whole syncytium (T) vs. frequency of the electric current. We 
suppose $\epsilon_{r}$ to be constant in the range of frequencies 
considered. We have chosen $\alpha = 45^{\circ}$ in order to have a mean
trajectory of the cosmic particle in the syncytium.

Figure \ref{fig4} shows the behavior of the energy lost vs. intensity of 
the electric current when $\alpha = 45^{\circ}$. 
 
\begin{figure*}[htbp]
\centerline{\resizebox{9cm}{9cm}{
\includegraphics*[width=10cm]{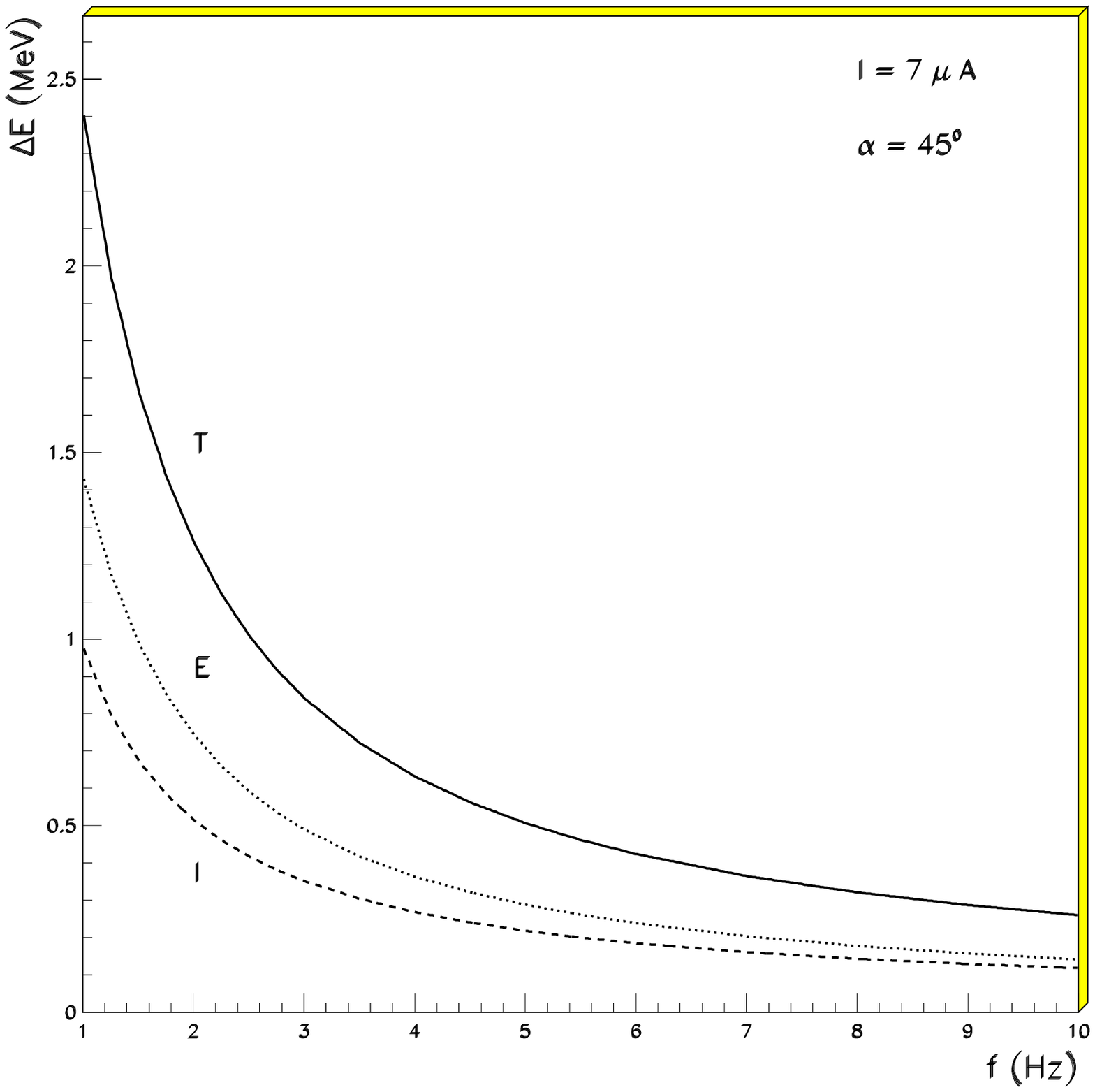}}}
\caption{\it{Energy lost in the syncytium vs. frequency.}} 
\label{fig3}
\end{figure*}

\begin{figure*}[htbp]
\centerline{\resizebox{9cm}{9cm}{
\includegraphics*[width=10cm]{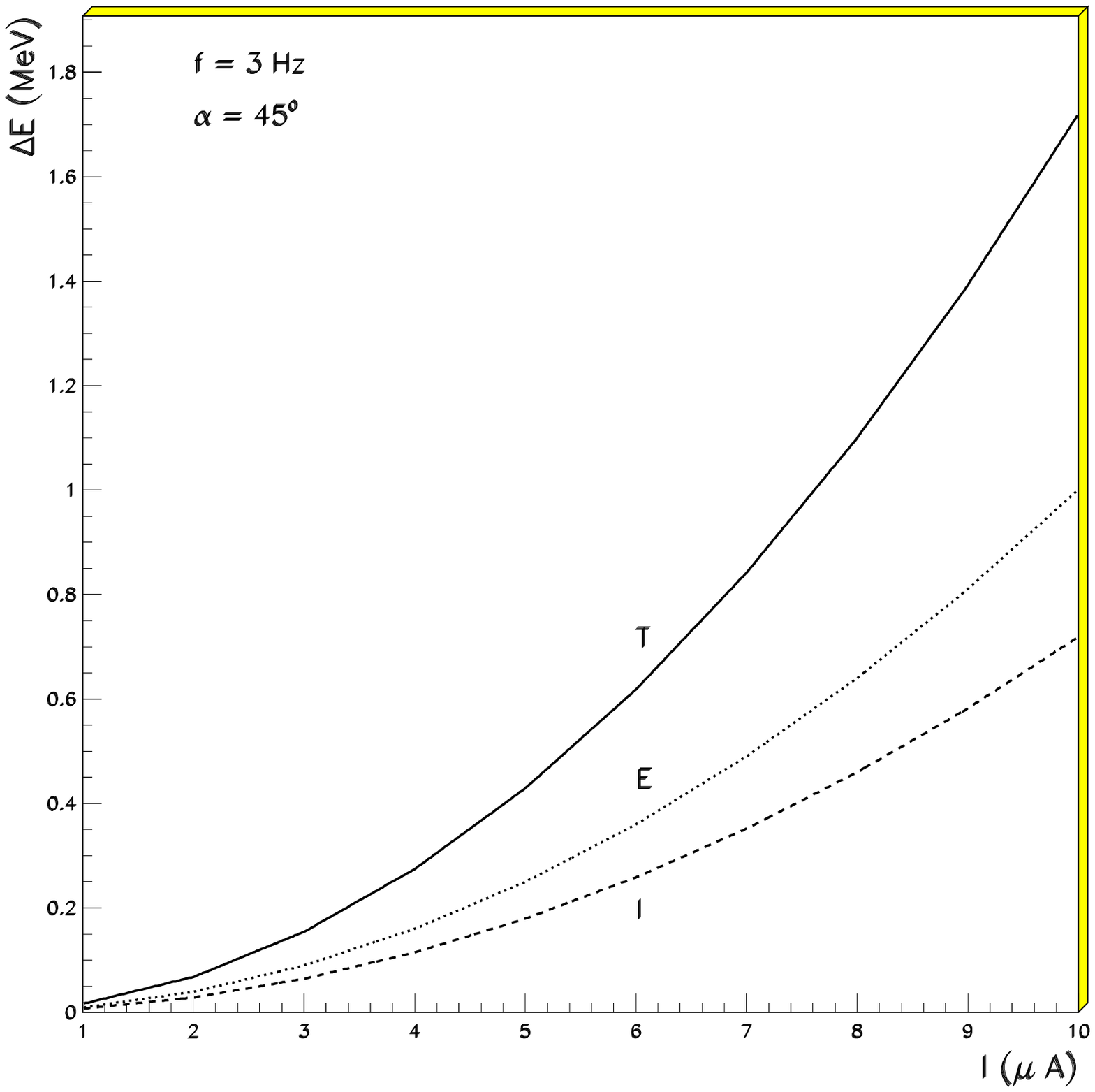}}}
\caption{\it{Energy lost in the syncytium vs. electric current.}} 
\label{fig4}
\end{figure*}

In Figure \ref{fig5} the most important result of this paper is shown, 
that is the energy lost in the syncytium vs. incidence angle $\alpha$ of
the cosmic charged particle. These are symmetric curves respect to the 
line $\alpha = 90^{\circ}$ because the system is symmetric with respect 
to the $z$ axis. We can see that the energy lost reaches the maximum 
value when the incidence angle of the cosmic charged particles is 
$180^{\circ}$. The minimum value is reached for $\alpha = 90^{\circ}$. 
Based on this fact we can suppose that the LF phenomenon occurs when 
cosmic charged particles pass through the astronaut visual system with a 
incidence angle of approximately $180^{\circ}$.  
 
\begin{figure*}[htbp]
\centerline{\resizebox{9cm}{9cm}{
\includegraphics*[width=10cm]{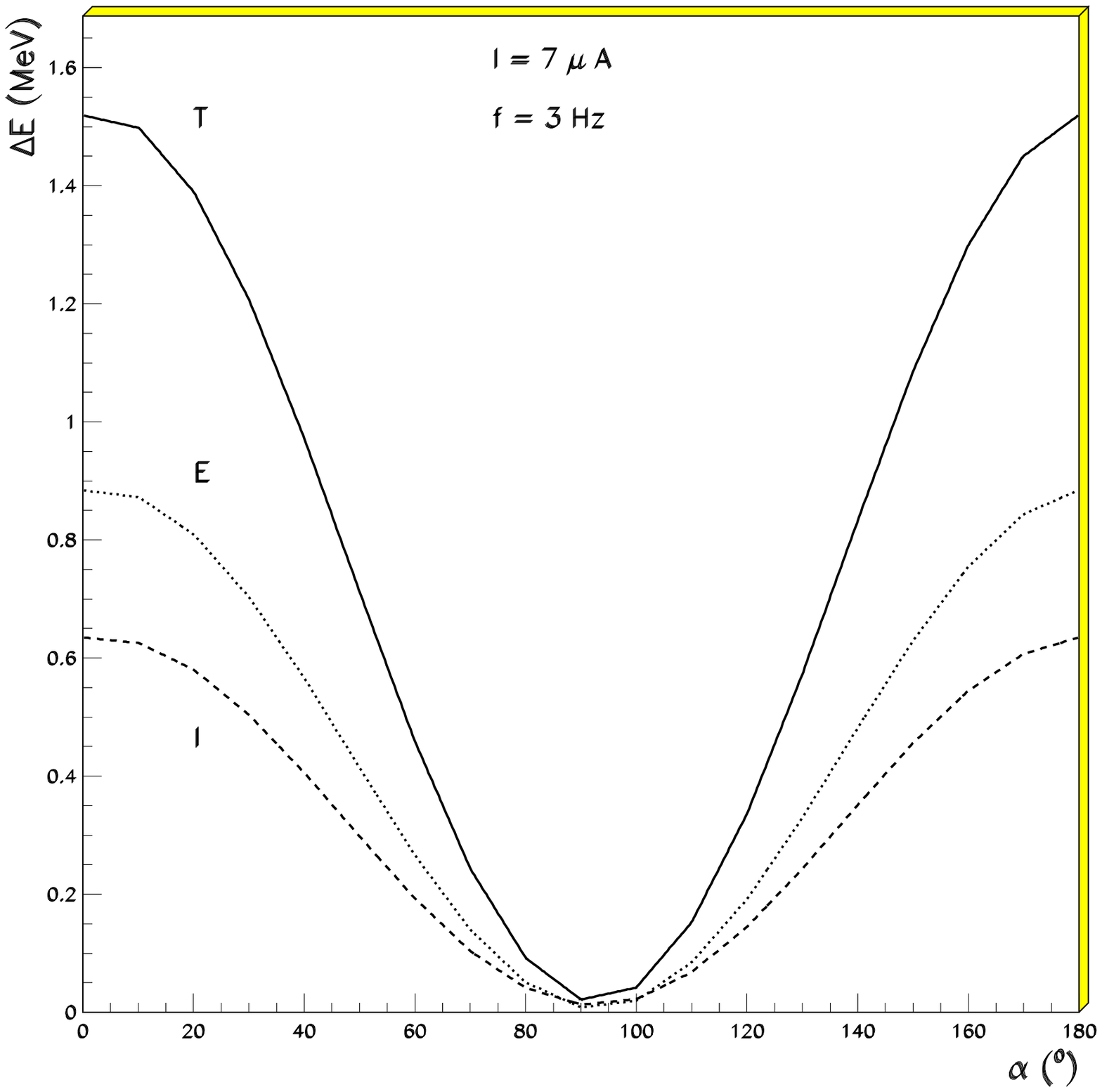}}}
\caption{\it{Energy lost in the syncytium vs. incidence 
angle of cosmic charged particle.}} 
\label{fig5}
\end{figure*}

\section*{\small{\centerline{5 INTERACTION ANTIMATTER-SYNCYTIUM}}}

When we consider the interaction between antimatter and syncytium, the 
right hand side of eq.(\ref{eq2}) must be changed because of the possible
annihilation process. We can write:

\begin{eqnarray}
\nabla \cdot [\epsilon_{r}(\underline{x}) \nabla U^{(i)}(\underline{x})]
+ R_{i} \alpha_{m} Y_{m} [U^{(e)}(\underline{x}) - 
U^{(i)}(\underline{x})] = -I R_{i} {{\partial \delta} \over 
{\partial \underline{v}_{I}}} (\underline{x} - \underline{x}_{I}) +
k \delta(\underline{x} - \underline{x}_{I})
\label{eq9}
\end{eqnarray}
with 
\begin{eqnarray}
k = 2 m c^{2} g 
\label{eq10}
\end{eqnarray}
where $m$ is the mass of the antiparticle and $g$ is a constant depending
on the antiparticle expressed in V/N. This constant will be determined 
experimentally in space. It is linked to cross section of the 
annihilation process. When $g = 0$ we are considering matter; when 
$g = 1$ we suppose that all the antiparticles entering in the syncytium
undergo the annihilation phenomenon. The true value of $g$ is much 
smaller than 1 since the antiparticles that we consider are flying.

In order to study the antimatter effect in the syncytium, we have chosen
the most favourable case ($g = 1$). We have seen that the effect of the 
annihilation process on the value of energy lost in the syncytium is 
negligible. In fact the source term 
$k \delta(\underline{x} - \underline{x}_{I})$ in eq.(\ref{eq9}) is much 
smaller than the other term when the current intensity $I$ is of order of
$\mu$A (see section 7, Table \ref{tab2}).

\section*{\small{\centerline{6 THE EVALUATION OF THE COSMIC RAY FLUXES}}}

We used the Creme96 computer program to evaluate the cosmic ray fluxes 
within the International Space Station. Creme96 is a package of computer
programs to create numerical models of the ionizing radiation 
environment in near Earth orbits and to evaluate the resulting radiation 
effects on electronic systems in spacecrafts and in high altitude flying 
aircrafts \it{(17-19)}\rm.

The differential fluxes, in minimum solar condition, are shown in 
Figure \ref{fig6}. There is a strong predominance of protons and alpha 
particles with respect to heavier nuclei and a great predominance of 
protons with respect to alpha particles when the kinetic energy is below
1 GeV/nucl. The maximum value for protons flux is near $10^2$ MeV/nucl. 
The remaining particles have maximum values near $10^3$ MeV/nucl.

\begin{figure*}[htbp]
\centerline{\resizebox{9cm}{9cm}{
\includegraphics*[width=10cm]{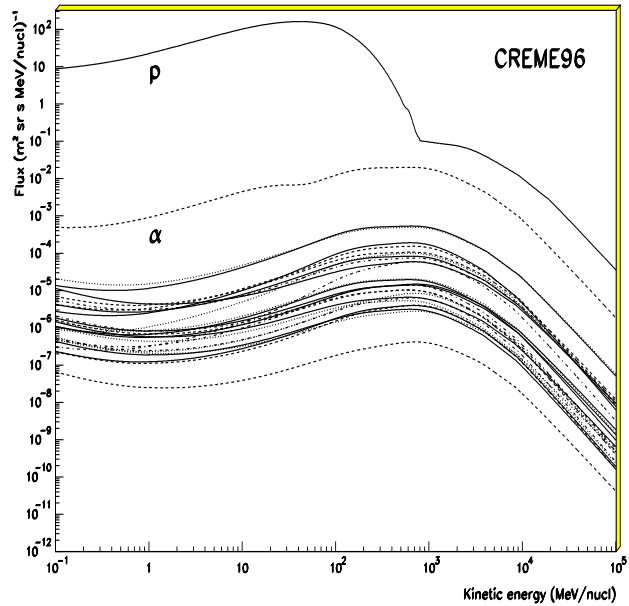}}}
\caption{\it{Cosmic ray differential fluxes vs. kinetic energy 
within the ISS.}}
\label{fig6}
\end{figure*}

\section*{\small{\centerline{7 THE GEANT SIMULATION}}}

In order to control the reliability of the syncytium model, we have 
developed a simulation with the Geant 3.21 program.
 
The Geant 3.21 program simulates the passage of elementary particles 
through matter. This program originally designed for High Energy Physics
experiments (HEP), today it has found applications also outside this 
domain in areas such as medical and biological sciences, radioprotection 
and astronautics.

The principal applications of Geant in HEP are:

(1) the transport of particles through an experimental setup for the
simulation of the detector response;

(2) the graphical representation of the setup and of the particle 
trajectories.

These two functions are combined in the interactive version of Geant. 
This is very useful, since the direct observation of what is happening 
to a particle inside the detector makes the debugging easier and may 
reveal possible weakness of the setup.

The Geant 3.21 program system can be obtained from CERN, European 
Organization for Nuclear Research, in 
http://cernlib.web.cern.ch/cernlib/version.html 
and the program runs everywhere the CERN Program Library is installed.

We remember that in \it{(6)} \rm at least two causes of Light Flashes are
hypothesized, one due to protons and the other due to heavy nuclei. For 
this reason we have developed a lot of simulations using the Geant 3.21 
program.  

In our simulation the incident particles are generated in a random way 
and isotropically on a big spherical surface with the crystalline lens 
in the centre. The lens is represented by a uniform sphere of water with 
the ray equal to $a$ because the Geant 3.21 program can not simulate an 
ellipsoid. The energy of a particle is chosen in a random way within the 
cosmic ray spectrum obtained using the Creme96 program 
(Figure \ref{fig6}), so that events are distributed according to this 
spectrum. The direction of a particle is isotropically generated in a 
random way. 

Figure \ref{fig7} shows events distribution vs. energy lost for cosmic 
protons that hit a sphere of water obtained by using the Geant 3.21 
program.

\begin{figure*}[htbp]
\centerline{\resizebox{9cm}{9cm}{
\includegraphics*[width=10cm]{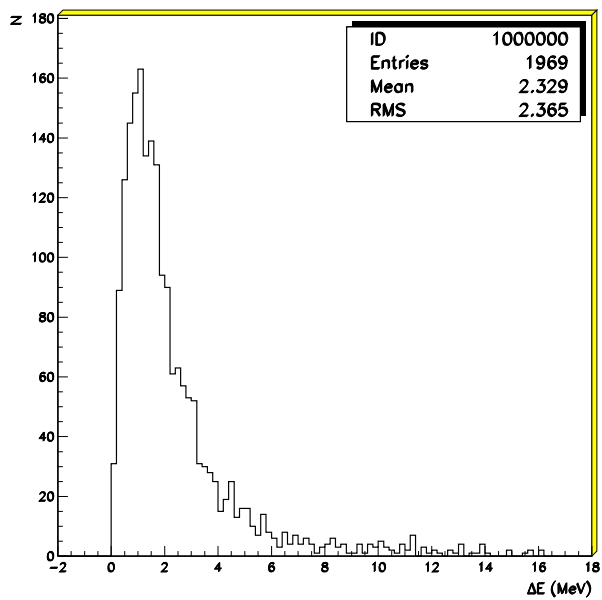}}}
\caption{\it{Events distribution vs. energy lost when a cosmic proton 
hits a sphere of water. The energy is expressed in MeV.}}
\label{fig7}
\end{figure*}

Table \ref{tab2} shows the comparison between the Geant simulation 
and the syncytium model. We remember that the sphere has radius $a$ and 
the ellipsoid has semiaxes $a$ and $a/2$. In the first column we report 
the kind of particle considered in the Geant simulations. The second 
column shows the average energy lost in a sphere of water of radius $a$ 
calculated by Geant. In the third column there is the electrical current
that in a uniform spherical syncytium with $\epsilon_{r}$ = 1 yield a 
lost of energy (fourth column) almost equal to the one calculated by 
Geant. The fifth, sixth and seventh columns show the energy lost in the 
intracellular and extracellular compartments and in the whole ellipsoidal
syncytium respectively having the parameters reported in Table \ref{tab1}
and the variable density described in section 3. In the last column we 
show the ratio between the energies lost in the spherical and ellipsoidal
syncytia. We can see that the energy lost in an uniform spherical 
syncytium is almost three times greater then the energy lost in the 
ellipsoidal syncytium having variable relative dielectric constant 
described previously. This difference is due to the difference in volume 
and in the dielectric constant $\epsilon_{r}$. 

The comparison between the Geant simulation and the syncytium model
indicates the reliability of the latter in the study of some electrical 
properties of the eye. 

\begin{table*}[htbp]
\vspace{1 cm}
\begin{center}
\begin{tabular}{|c|c|c|c|c|c|c|c|}
\hline
Particle   &   $\Delta E_{G}$   &   $I$   &   $\Delta E_{s}$   &  
$\Delta E^{I}_{e}$   &   $\Delta E^{E}_{e}$   
&   $\Delta E^{T}_{e}$   &   $\Delta E_{s}/\Delta E^{T}_{e}$  \\   
\hline
p    &   2.329    &    7    &   2.329    &   0.351
&   0.490     &    0.841    &   2.77    \\
\hline
$\alpha$      &    1.833    &      6     &    1.909
&    0.258    &    0.360    &    0.618    &   3.09     \\
\hline
$^{7}\mbox{Li}_3$    &    4.044    &      9     &    4.295
&    0.581    &   0.809     &    1.390    &   3.09     \\
\hline
$^{10}\mbox{B}_5$    &    11.54    &     15     &    11.93
&    1.615    &    2.249    &    3.864    &   3.09     \\
\hline
$^{12}\mbox{C}_6$    &    16.46    &     17     &    15.32
&    2.074    &    2.888    &    4.962    &   3.09     \\
\hline
$^{14}\mbox{N}_7$    &    22.21    &     20     &    21.21
&    2.871    &    3.998    &    6.869    &   3.09     \\
\hline
$^{16}\mbox{O}_{8}$  &    28.80    &     23     &    28.05
&    3.801    &    5.286    &    9.087    &   3.09     \\
\hline
$^{56}\mbox{Fe}_{26}$  &  308.2    &     76     &    306.3
&    41.50    &    57.72      &  99.22    &   3.09     \\
\hline
\end{tabular}
\end{center}
\caption{\it Comparison between the syncytium model and the Geant 
simulation. The energy is expressed in MeV. The electric current is
expressed in $\mu$A. The values of the parameters are f = 3 Hz and 
$\alpha = 45^{\circ}$.}
\label{tab2}
\end{table*}

\section*{\small{\centerline{8 THE SPACE RADIATION EFFECTS}}}

The health risk to astronauts from cosmic rays radiation determines the
maximum lenght of space missions. As a consequence it is very important 
to evaluate the effects of charged particles on organs of the human body.
It is necessary to have a set of dosimetric codes to convert the 
radiation environment within spacecrafts into radiation protection 
quantities, which can be used to evaluate astronaut risk when exposure 
limits have been established. These limits exist for Low Earth Orbit 
(LEO) only. For missions beyond the protection of the Earth's magnetic 
field, risk increases. In each case the shielding of spacecrafts is 
basic.

We studied the effects of cosmic radiation on an ``eye'' simulated by 
Geant 3.21 program. 

We remember some definitions. The \it{absorbed dose} \rm $D_{1}$ is the 
quantity which measures the total energy absorbed per unit mass and is 
the fundamental parameter in radiological protection. Then we have:

\begin{eqnarray}
D_{1} = {\Delta E_{G} \over M_{G}},
\label{eq11}
\end{eqnarray}
where $M_{G}$ is the mass of the uniform sphere of water simulated by 
Geant. The unit of $D_{1}$ is the Gray which is defined as 1 Gy = 1 J/kg.

The \it{absorbed dose} \rm describes the physical effect of the incident 
radiation, but it gives no information on the rate of absorption and on 
the specific type of the radiation. These factors are very important when
considering the biological effects of radiation, then $D_{1}$ is an 
inadequate quantity. For example, an absorbed dose of $\alpha$ particles 
produces more damage than an equal dose of protons, and a given dose of 
protons produces more damage than a similar dose of electrons or 
$\gamma$-rays. In fact different particles deposit locally different 
energy per unit path lenght. Thus the particles with bigger ionizing 
power yield a greater local biological damage.

For considering this effect, to each radiation type is assigned a 
radiation weighting factor, $w$ \it{(20)}\rm. The factors are independent
from tissue type, are experimentally determined and have stochastic 
character. The quality factor of a radiation type is defined as the ratio
between the biological damage produced by the absorption of 1 Gy of that 
type of radiation and the biological damage produced by 1 Gy of X or 
$\gamma$ radiation.

Then the \it{equivalent dose}, \rm $H$, is obtained multiplying the 
value of the \it{absorbed dose}, \rm averaged over the entire tissue or 
organ, by the radiation weighting factor: 

\begin{eqnarray}
H = D_{1} \times w.
\label{eq12}
\end{eqnarray}
The \it{equivalent dose} \rm expresses long-term risk (primarily cancer 
and leukemia) from low-level chronic exposure.

The unit of \it{equivalent dose} \rm is the Sievert (Sv) which has the 
same dimensions as the Gray (J/kg), but now 1 Sv of $\alpha$ particles 
produces approximately the same effect as 1 Sv of X or $\gamma$-rays, 
etc. Howewer the \it{equivalent dose} \rm is not a quantity directly 
measurable while the \it{absorbed dose} \rm is directly observable.
 
If more than one radiation type is present, the total biological effect 
suffered by a tissue or organ is:

\begin{eqnarray}
H_{tot} = \sum_{R} D_{R} w_{R},
\label{eq13}
\end{eqnarray}
where $D_{R}$ is the average \it{absorbed dose} \rm received by the organ
from the radiation type ${R}$ having a weighting factor equal to $w_{R}$.

Table \ref{tab3} shows the \it{absorbed} \rm and the \it{equivalent 
dose} \rm in the crystalline lens (uniform sphere of water) relative to 
the average energy lost calculated by the Geant simulation. Then we 
obtain $H_{tot} = 74.575 \ \mu$Sv.  

\begin{table*}[htbp]
\vspace{1 cm}
\begin{center}
\begin{tabular}{|c|c|c|c|c|}
\hline
Particle   &   $\Delta E_{G}$ [MeV]    &     $D$ [$\mu$Gy]    &    $w$    &    $H$ [$\mu$Sv]  \\
\hline
p                     &    2.329    &   0.022    &     5    &    0.110  \\
\hline
$\alpha$              &    1.833    &   0.017    &    20    &    0.347  \\
\hline
$^{7}\mbox{Li}_3$     &    4.044    &   0.038    &    20    &    0.766  \\
\hline
$^{10}\mbox{B}_5$     &    11.54    &   0.109    &    20    &    2.186  \\
\hline
$^{12}\mbox{C}_6$     &    16.46    &   0.156    &    20    &    3.118  \\
\hline
$^{14}\mbox{N}_7$     &    22.21    &   0.210    &    20    &    4.207  \\
\hline
$^{16}\mbox{O}_{8}$   &    28.80    &   0.273    &    20    &    5.456  \\
\hline
$^{56}\mbox{Fe}_{26}$ &    308.2    &   2.919    &    20    &   58.385  \\
\hline
\end{tabular}
\end{center}
\caption{\it Absorbed and equivalent energy doses in the crystalline lens
corresponding to the average energy lost calculated by Geant simulation. 
The energy is expressed in MeV. Absorbed and equivalent energy doses are 
expressed in $\mu$Gy and in $\mu$Sv respectively.}
\label{tab3}
\end{table*}

Table \ref{tab4} shows the \it{absorbed} \rm and the \it{equivalent 
dose} \rm in the crystalline lens relative to the average energy lost in
one year calculated by the Geant simulation. Here $q$ is the interactions
number per second occurring in the lens and 
$\Delta E_{G}^{s} = \Delta E_{G} \times q$ is the mean energy lost in 
one second. Without protons we obtain $H_{tot} = 96.4$ mSv/yr.  

\begin{table*}[htbp]
\vspace{1 cm}
\begin{center}
\begin{tabular}{|c|c|c|c|c|c|}
\hline
Particle                &            $q$               &    $\Delta E_{G}$ [MeV]    &    $\Delta E_{G}^{s}$ [eV]   &         $D$ [mGy/yr]         &         $H$ [mSv/yr]      \\
\hline 
p                       &           1.508              &            2.329           &     $3.51 \times 10^{6}$     &     $1.048 \times 10^{3}$    &    $5.24 \times 10^{3}$   \\
\hline
$\alpha$                &    $3.79 \times 10^{-3}$     &            1.833           &            6950              &             2.08             &             41.6          \\
\hline        
$^{7}\mbox{Li}_3$       &    $1.83 \times 10^{-5}$     &            4.044           &            74.0              &     $2.21 \times 10^{-2}$    &            0.442          \\
\hline
$^{10}\mbox{B}_5$       &    $3.03 \times 10^{-5}$     &            11.54           &             349              &    $10.45 \times 10^{-2}$    &              2.1          \\
\hline
$^{12}\mbox{C}_6$       &    $10.2 \times 10^{-5}$     &            16.46           &            1679              &    $50.15 \times 10^{-2}$    &               10          \\
\hline
$^{14}\mbox{N}_7$       &    $2.71 \times 10^{-5}$     &            22.21           &             602              &    $17.98 \times 10^{-2}$    &              3.6          \\
\hline
$^{16}\mbox{O}_{8}$     &    $9.76 \times 10^{-5}$     &            28.80           &            2811              &    $83.97 \times 10^{-2}$    &             16.8          \\
\hline        
$^{56}\mbox{Fe}_{26}$   &    $1.19 \times 10^{-5}$     &            308.2           &            3668              &    $109.6 \times 10^{-2}$    &             21.9          \\
\hline
\end{tabular}
\end{center}
\caption{\it Absorbed and equivalent energy doses in the crystalline lens
corresponding to the average energy lost in 1 year calculated by Geant 
simulation.}
\label{tab4}
\end{table*}

In order to give an idea of character of the numbers of Table \ref{tab4},
we cite the dose limits as recommended by the International Commission on
Radiological Protection (ICRP) \it{(20)}\rm. Two sets of limits are 
defined: one for individual exposed occupationally and one for the 
general public. For the lens of eye the limits are 150 mSv/yr 
(occupational) and 15 mSv/yr (general public). These are allowable doses 
in addition to the natural background dose. In radiotherapy the doses 
given to the tumour are typically around 100 to 200 Sv per treatment.

Examples of use of codes for calculating the dosimetric quantities for 
several near Earth environments can be found in \it{(21)}\rm.

Results of measurements of the absorbed and equivalent dose on board  
aircrafts, spacecrafts and space station Mir can be found in \it{(22)}\rm. A
discussion of the planned radiation measurement on the International 
Space Station is given in \it{(23)}\rm.

\section*{\small{\centerline{9 CONCLUSIONS}}}
The comparison with the results obtained with the Geant simulation 
program shows that the modellization of part of the human visual system 
with an ellipsoidal syncytium is promising. The work presented in this 
paper suggests that this model can be used in the qualitative study of 
some unexpected phenomena such as the Light Flashes observed by the 
astronauts.
\newpage

\section*{\small{\centerline{REFERENCES}}}
\noindent

\it{1}. \rm W.Z. Osborne, L.S. Pinsky,  J.V. Bailey, Apollo light 
flashes investigations. In: Johnston R.S., Dietlein L.F., Berry C.A. 
(Eds.), \it{Biomed. results of Apollo}\rm. NASA-STIO 355-365 (1975).

\it{2}. \rm L.S. Pinsky, W.Z. Osborne, J.V. Bailey, R.E. Benson, 
L.F. Thompson, Light flashes observed by astronauts on Apollo 11 
through Apollo 17. \it{Science} \bf{183}\rm, 957-959 (1974).

\it{3}. \rm L.S. Pinsky, W.Z. Osborne, R.A. Hoffman, J.V. Baily, 
Light flashes observed by astronauts on Skylab 4. 
\it{Science} \bf{188}\rm, 928-930 (1975).

\it{4}. \rm M. Casolino, V. Bidoli, E. De Grandis, M.P. De Pascale, 
G. Furano, A. Morselli, L. Narici, P. Picozza, E. Reali, C. Fuglesang, 
Study of the radiation environment on MIR space station with 
Sileye-2 experiment. \it{Adv. Space Res.} \bf{31}\rm, 1, 135-140 (2003).

\it{5}. \rm M. Casolino, V. Bidoli, G. Furano, M. Minori, A. Morselli, 
L. Narici, P. Picozza, E. Reali, R. Sparvoli, P. Spillantini,
The Sileye-3/Alteino experiment on board International Space 
Station. \it{Nucl. Phys. B (Proc. Suppl.)} \bf{113}\rm, 71-78 (2002). 

\it{6}. \rm M. Casolino, V. Bidoli, A. Morselli, L. Narici, 
M.P. De Pascale, P. Picozza, E. Reali, R. Sparvoli, G. Mazzenga, 
C. Fuglesang, Dual origins of light flashes seen in space. 
\it{Nature} \bf{422}\rm, 680 (2003).

\it{7}. \rm L. Narici, V. Bidoli, M. Casolino, M.P. De Pascale, 
G. Furano, I. Modena, A. Morselli, P. Picozza, E. Reali, C. Fuglesang, 
The ALTEA facility on the International Space Station. 
\it{Phys. Med.} \bf{17}\rm, Supplement 1, 255-257 (2001).

\it{8}. \rm L. Narici, V. Bidoli, M. Casolino, M.P. De Pascale, 
G. Furano, A. Iannucci, A. Morselli, P. Picozza, E. Reali, C. Fuglesang, 
ALTEA: Visual perception studies on astronauts on board the 
ISS. \it{Proc. of ICRC 2001}\rm, 2322-2323, Copernicus Gesellschaft 
2001.

\it{9}. \rm L. Narici, V. Cotronei, P. Picozza, W.G. Sannita, A. Galper,
V.P. Petrov, V.P. Salnitskii, ALTEA: Investigating the effect of 
particles on human brain functions on ISS. \it{AIAA}\rm, Paper 2001-4942
(2001).

\it{10}. \rm L. Narici, V. Bidoli, M. Casolino, M.P. De Pascale, 
G. Furano, A. Morselli, P. Picozza, E. Reali, R. Sparvoli, C. Fuglesang, 
ALTEA: Anomalous long term effects in astronauts. A probe on the 
influence of cosmic radiation and microgravity on the central nervous 
system during long flights. \it{Adv. Space Res.} \bf{31}\rm, 1, 141-146 
(2003).

\it{11}. \rm J.L. Rae, The electrophysiology of the crystalline 
lens. In: \it{Curr. Top. in Eye Res.} \bf{1}\rm, 37-90 (1979).

\it{12}. \rm R.S. Eisenberg, V. Barcilon, R.T. Mathias, Electrical 
properties of spherical syncytia. \it{Bioph. J.} \bf{25}\rm, 151-180 (1979). 

\it{13}. \rm R.T. Mathias, J.L. Rae, R.S. Eisenberg, The lens as a 
nonuniform spherical syncytium. \it{Biophys. J.} \bf{34}\rm, 61-83 (1981). 

\it{14}. \rm P. Maponi, M. Ricci, B. Spataro, F. Zirilli, A syncytium
model for the interpretation of the phenomenon of anomalous light 
flashes occuring in the human eye during space missions. 
\it{Nuovo Cimento Soc. Ital. Fis., B} \bf{116}\rm, 1173-1179 (2001).

\it{15}. \rm E.N. Pugh Jr., B. Falsini, A.L. Lyubarsky, The origin 
of the major rod-and-cone-driven components of the rodent 
electroretinogram and the effect of age and light-rearing history on the 
magnitude of these components. \it{Photostatis and Related Phenomena}\rm, 
93-128. Williams and Thistle Editors, Plenum Press, New York, 1998.

\it{16}. \rm G.H. Golub, C.F. Van Loan, \it{Matrix Computation}\rm, 
3rd Edition, The Johns Hopkins University Press, Baltimore, US, 1996.

\it{17}. \rm A.J. Tylka, W.F. Dietrich, P.R. Boberg, E.C. Smith, 
J.H. Adams, Jr., Single Event Upsets Caused by Solar Energetic Heavy 
Ions. \it{IEEE Trans. on Nucl. Sci.} \bf{43}\rm, 2758-2766 (1996).

\it{18}. \rm A.J. Tylka, J.H. Adams Jr., P.R. Boberg, B. Brownstein, 
W.F. Dietrich, E.O. Flueckiger, E.L. Petersen, M.A. Shea, D.F. Smart, 
E.C. Smith, CREME96: A Revision of the Cosmic Ray Effects on 
Micro-Electronics Code. \it{IEEE Trans. on Nucl. Sci.} \rm 
\bf{44}\rm, 2150-2160 (1997); and references therein.

\it{19}. \rm A.J. Tylka, W. F. Dietrich, P.R. Boberg, Probability 
Distributions of High-Energy Solar-Heavy-Ion Fluxes from IMP-8: 
1973-1996. \it{IEEE Trans. on Nucl. Sci.} \bf{44}\rm, 2140-2149 (1997).

\it{20}. \rm 1990 Recommendations of the International Commission for
Radiological Protection. \it{ICRP Rep.} \rm No. 60, Annuals of the 
ICRP 21, No. 1-3, Elsevier Science, New York, 1991.

\it{21}. \rm M.S. Clowdsley, J.W. Wilson, M.H. Kim, B.M. Anderson, 
J.E. Nealy, Radiation Protection Quantities for Near Earth 
Environments. \it{AIAA} \rm Space 2004 Conference and Exposition, 
San Diego, California, September 28-30 (2004).

\it{22}. \rm http://www.ati.ac.at/$\sim$vanaweb/publications.html.

\it{23}. \rm G.D. Badhwar, Radiation Measurement on the International 
Space Station. \it{Phys. Med.} \bf{17}\rm, Supplement 1, 287-291 
(2001).

\end{document}